# Monitoring of structural changes in materials under the exposure of ionization radiation using a vibrating wire

S.G. Arutunian[a,b*], N.M. Manukyan[a], S.A. Hunanyan[a], A.V. Margaryan[a], E.G. Lazareva[a*], M.Chung[c], L.M. Lazarev[a], G.S. Harutyunyan[a], D.A. Poghosyan[a], N.B. Margaryan[a] N.R. Aghamalyan[d], M.N. Nersisyan[d]

[a] *Alikhanyan National Scientific Laboratory Yerevan, 0036, Armenia*

[b] *CANDLE Institute for Synchrotron Research, Yerevan, 0022, Armenia*

[c] *Pohang University of Science and Technology, Pohang, Gyeongbuk, 37673, South Korea*

[d] *Institute for Physical Research of National Academy of Sciences of Armenia, Ashtarak, 0204, Armenia*

[*] *Corresponding author: S.G. Arutunian: (e-mail: femto@yerphi.am)*

Keywords: irradiation, proton beam, X-ray, vibrating wire, structural changes, electrical pulses

**Abstract**

Ionizing radiation (X-rays, proton beams) causes structural changes in materials. If a vibrating metallic wire is subjected to such radiation, the natural frequency of the wire is affected as a result of changes in the elastic characteristics of the material. This paper presents the results of experiments on the impact of X-ray radiation in the range of 100-165 keV and a proton beam with energy 18 MeV on the structure of stainless steel wire. In case of proton irradiation an irreversible change in the wire frequency was observed, which indicated residual changes in the structure of the wire material. X-ray diffractometry methods were used to analyze the structural changes.

## 1. Introduction

The operating principle of vibrating wire monitors (VWM) is based on dependence of natural frequency of the stretched on a support wire on the physical parameters of the wire and environment in which oscillations take place. The area of application of this technique is wide and the number of vibrating wire-based instruments has increased. Strain (Bachofner et al, 2023), displacement (Vibrating Wire Displacement Transducers, 2025), piezometric level (Vibrating Wire Piezometer – Types and Operating Principle, 2025), pressure (Wan et al, 2013), angle (Vibrating wire tiltmeter, 2025), viscosity of the media (Richter et al, 2016), and ultralow thermometry (Zavjalov, 2023) are measured by vibrating wire instruments.

The important advantages of properly constructed vibrating wire sensors are inherent long-term stability (Simmonds, 2013), high precision and resolution, good reproducibility and small hysteresis. The advantage of vibrating wire sensors is also that the frequency signal can be transmitted over long cables with no loss or degradation of the signal. The reliability of the sensors becomes the overriding feature in the selection of a technology. It is also is important to ascertain a low drift and minimum change in sensitivity. An important parameter of vibrating wire-based sensors is their ability to operate in hard conditions (high operational power and temperature cycling, thermal shock, thermal storage, autoclaving, fluid immersion, mechanical shock, electromagnetic and electrostatic environments) (Agejkin, 1965; Asch, 1991; Bourquin et al, 2005; Jing et al., 2024).

Vibrating wire monitors (VWM), developed by the accelerator diagnostics group at Yerevan Physics Institute, have been used for many years, mainly for profiling various accelerator beams, as well as photon beams with a wide energy spectrum (Arutunian et al., 2021). It appears that vibrating wire monitors are a good addition to the traditional thin wire scanning method, in which the signal is created by measurement of a flux of secondary radiation/particles (Wittenburg, 2013; Igarashi et al., 2002; Nazhmudinov et al., 2018; Nazhmudinov et al., 2022). VWM are more sensitive and can be used to measure beam halos, however, have lower operation speed, as they require wire thermalization during scanning.

The principle of VWM for accelerated diagnostics is based on the strong dependence of the natural frequency of a wire on its temperature when the ends of the wire are rigidly fixed and the wire is exposed to a heat source (beams of charged particles in accelerators, neutron and photon beams). Short electrical pulses can also be used as heat sources (for more details, see section 3). The monitors we developed used the ability to continuously generate wire vibrations at its natural frequency. The basis was taken from an electrical scheme (Eltsev et al. 1984), in which the wire is connected in a positive feedback circuit of an operational amplifier (see also (Arutunian et al. 2007)). The mechanical effect of the current on the wire was due to the presence of a magnetic system surrounding the wire in certain areas (position 4 in Fig. 1). The circuit was subsequently significantly upgraded by I. Vasiniuk (Arutunian et al. 2014): the effect of pulses feeding the wire's oscillations on the process of the wire's oscillations at its natural frequency was minimized, and the stabilization of the oscillation amplitude was significantly improved. The frequency resolution for VWM corresponds to 0.01 Hz for 1 s measurements which corresponds to thermal resolution 0.3 mK for stainles steel, 0.6 mK for berillium bronze, and 1 mK for tungsten (Arutunian et al. 2021).

The main type of single wire VWM modifications is shown in Fig. 1. The wire material can be heat-treated stainless steel (hardened by heating to a critical temperature, rapid cooling (quenching), and then tempering), beryllium bronze or tungsten. After preliminary tensioning, the wire is fixed in clamps. The monitors maintain their working tension for many years. The initial frequency of the VWM depends on the length and diameter of the wire, as well as the initial tension of the wire material, and is in the range of 1-10 kHz.

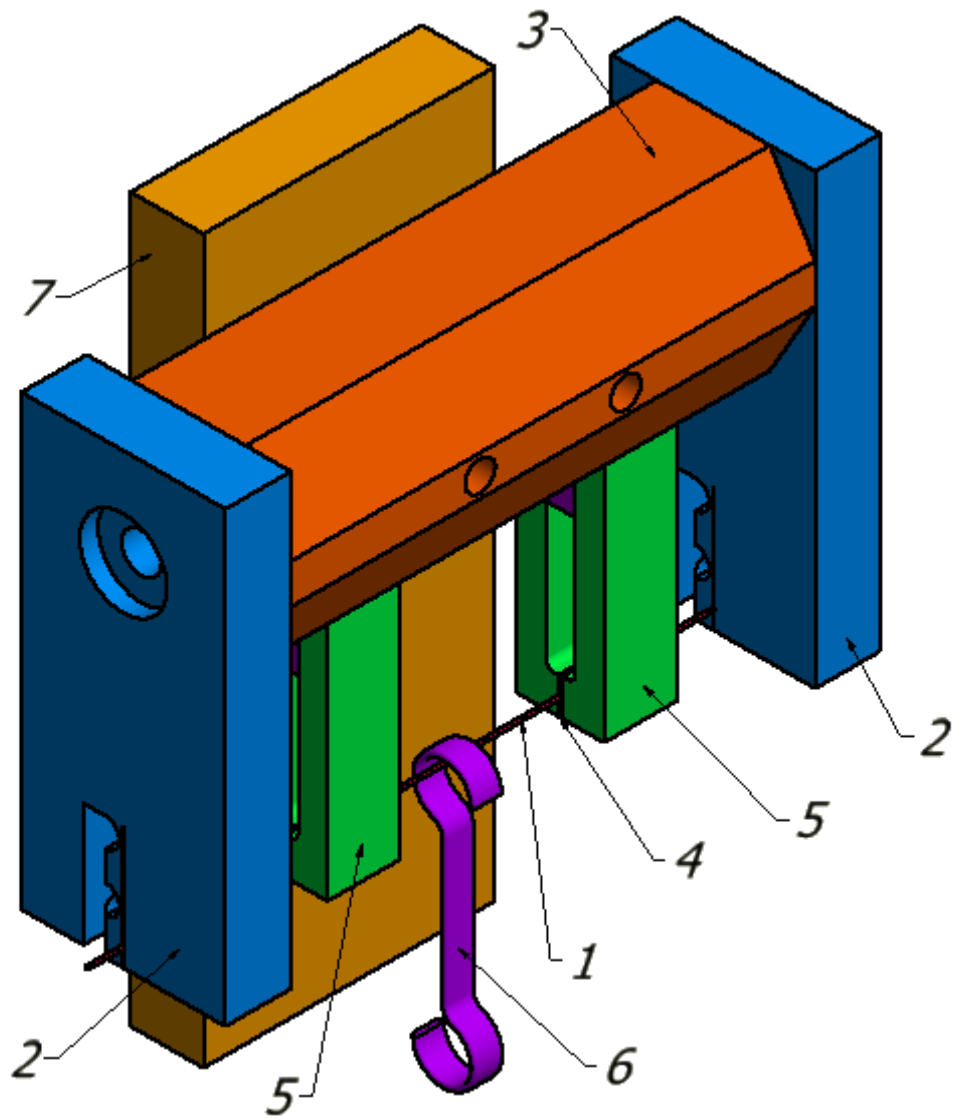

Figure 1. General view of a single wire sensor: 1 – vibrating wire, 2 – clips clamping the wire, 3 – VWM base, 4 – gaps in the magnetic system (5) based on permanent magnets, 6 - hook, 7 - support rod.

The profiling station based on two VWM (for horizontal and vertical beam scanning), which was originally used to measure the profile of the C18 cyclotron proton beam (Aginian et al., 2021; Arutunian et al., 2024a), was also used to study structural changes in wire material subjected to proton irradiation.

## 2. Wire tension above the elastic limit and application of short electrical pulses

Under laboratory conditions, structural changes in materials can be modeled using two procedures that are simple to perform experimentally: subjecting the material to loads above its elastic limit and applying thermal treatment.

The paper (Arutunian et al., 2024b; Arutunian et al., 2024c) describes studies where a tension of the order of tensile stress is applied to a wire-shaped specimen so that the wire material undergoes plastic deformation. The wire was loaded as follows. The vibrating wire monitor was mounted on a support so that the wire was oriented horizontally (see Fig. 1). A flat hook (pos. 6 on Fig. 1) with smooth edges was attached to the middle of the wire, from which a basket with weights was suspended. The masses of the weights were increased. After each such procedure, the weights were removed from the wire (together with the hook) and the monitor switched to frequency generation mode. When a load is suspended, the tension in the wire increases, which can lead to plastic deformation and corresponding structural changes in the wire (see, for example, (Roters et al. 2010)). Let us determine this tension in the case when the wire is pre-tensioned between the clips. This pre-tension $\sigma_{VWM}$ is related to the elongation $\Delta L_{VWM}$ of the wire during its fastening in the clips by the equation

$$\Delta L_{VWM} = L \sigma_{VWM} / E, \tag{1}$$

where $L$ - is the length of the stretched wire (distance between clips), $E$ - is the modulus of elasticity of the wire material. The relation between $\sigma_{VWM}$ and the natural frequency $F_{VWM}$ of the wire in the monitor (generation at the second harmonic) is determined by the formula

$$\sigma_{VWM} = F_{VWM}^2 L^2 \rho, \tag{2}$$

where $\rho$ is density of wire material.

When a wire is loaded at its center by a vertically directed transverse force $P = mg$ ($m$ is the mass of the load on the wire, $g$ is the free-fall acceleration), the wire deflects at an angle $\theta$ relative to its direction and its length stretches by an additional to (1) value:

$$\Delta L_P = L(1/\cos\theta - 1). \tag{3}$$

At the same time, stress $\sigma_P$ in the wire (along the wire direction) increases according to the formula

$$\sigma_P = \sigma_{VWM} + E(1/\cos\theta - 1). \tag{4}$$

On the other hand, we have the condition of equilibrium for a load suspended from a wire:

$$2\sigma_P S \sin\theta = mg, \tag{5}$$

where $S$ is cross-section of the wire.

Solving (4) and (5) with respect to the angle $\theta$, we can find the equation that defines the stress $\sigma_P$ in the wire as a function of the mass $m$ loaded onto the wire and the value of the initial tension $\sigma_{VWM}$ in the wire:

$$\frac{4\sigma_P^4}{\sigma_m^2 E^2} + \frac{8\sigma_P^3 A}{\sigma_m^2 E} + \sigma_P^2 \left( \frac{4(A^2-1)}{\sigma_m^2} - \frac{1}{E^2} \right) - \frac{2\sigma_P A}{E} - A^2 = 0, \tag{6}$$

where $A = 1 - \sigma_{VWM} / E$, $\sigma_m = mg / S$.

Fig. 2 shows the results of the experiment on the wire. VWM parameters: wire from stainless steel (A316), $L$ = 72 mm, $d$ = 100 μm, generation of oscillations on second harmonics. For each load mass is given the value $\sigma_P$ calculated using formula (6), as well as the frequency of the wire $F_{VWM}$ after removing the load and corresponding stress in vibrating wire $\sigma_{VWM}$.

In the range of weights 0-50 g, the frequency of the wire's oscillations practically returned to its initial value. Starting from weights of 250 g, a sharp drop in frequency was observed. The wire broke at values greater than 350 g.

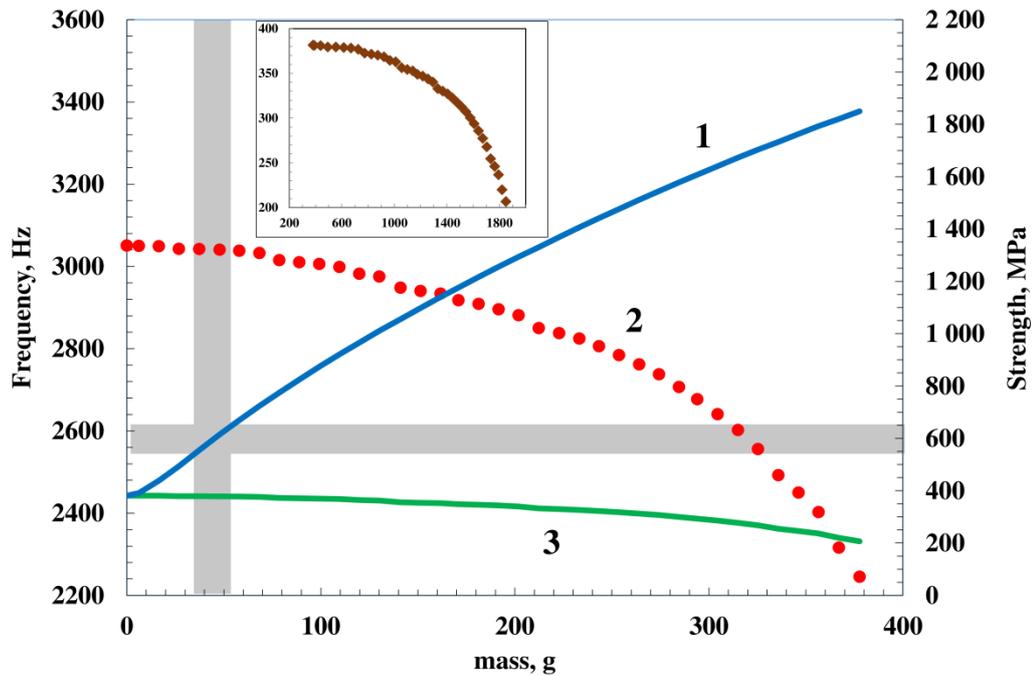

Figure 2. Experiments in which the load on the wire was gradually increased. 1(blue line) - stress values in the wire during the suspension of weights (calculated by Eq. 6). The horizontal gray strip indicates the elastic limit for the steel grade used (550-650 MPa). At stresses above this strip, residual deformations of the wire material are observed. 2(red circles) - frequency values after the weight was removed from the wire and the wire was connected to an autogenerator of natural oscillations (experimantal data). The vertical gray strip indicates changes in the monitor frequency after the load is removed in the frequency values to the right of this strip. 3(green line) - stress values in the wire corresponding to the frequency of natural oscillations (calculated by Eq. 2). The insert shows the dependence of the residual stress in the wire (vertical axis) on the stress created in the wire when a load is applied (horizontal axis). The values are given in MPa.

As can be seen in Fig. 2, the wire loading procedure leads to the occurrence of stresses exceeding the elastic limit (for the A316 steel grade used, in the order of 550-650 MPa) (Elastic Limit: Critical Threshold for Steel Performance & Design, 2025), which affects the magnitude of the residual frequency change after the load is removed

It is known that thermal treatment of materials (combination of heating and cooling operations) also leads to structural changes. We are particularly interested in procedures that lead to hardening (embrittlement), which in a sense are opposite to plastic deformation processes. We are most interested in quenching operations (see, for example, (Troell et al. 2024)), in which the sample is heated and then rapidly cooled, which leads to hardening of the material and, accordingly, to an increase in the elastic modulus of the wire material and an increase in the natural frequency. Research and monitoring of the embrittlement process is an actual task in nuclear power industry, where one of the structural materials of the reactor vessel is stainless steel, and the embrittlement process is one of the main factors determining the operating life of the reactor (Was 2011).

In the works (Arutunian et al., 2024b; Arutunian et al., 2024c), the hardening procedures were performed by applying short electrical pulses to a wire. The wire is rapidly overheated due to the short pulse duration and then cooled quite rapidly due to natural convection in the air.

A short electrical pulse was generated using a capacitor battery with variable charging voltage (battery capacity was 48 μF). To make the pulse as short as possible, a thyristor switch was used to short-circuit the pulse to the wire. Pulse parameters: pulse duration 250 μs at half-height, pulse amplitude up to 300 V, wire resistance about 10 Ohms. Thus, a current of several tens of amperes flowed through the string for a short time. After the pulse was applied to the wire, the wire was disconnected from the capacitor battery and connected to the autogenerator. Depending on the pulse amplitude, an increase in the natural frequency of the wire was observed, which we interpret as the wire hardening effect.

In the work (Arutunian et al., 2024c), the following experiment was carried out. A weight of 250 g was selected, as a load, which reduced the frequency, by approximately 100 Hz. Next, a short pulse with a voltage amplitude of 153 V (maximal value of current about 15 A) was applied to harden the wire. The experiment was conducted as follows. The initial frequency of the monitor was measured. Next, the wire was subjected to a load as described above. After this procedure, the frequency of the wire was measured for approximately two minutes. Next, the wire was disconnected from the self-oscillation circuit board, and the isolated ends of the wire were connected to the terminals of the capacitor bank. Next, a short electrical pulse was applied to the wire. The wire was then reconnected to the self-oscillation circuit and the frequency was measured for approximately two minutes. The process was repeated.

Several characteristic frequency jumps between corresponding levels (upper - after the hardening procedure, lower - after the wire loading procedure) are shown in Fig. 3. Linear regressions of these levels are also shown, demonstrating fairly good reproducibility of the processes over time. Linear regression equations: $F_{VWM}[Hz] = 0.00037 t[s] + 2849.9$ for upper level and $F_{VWM}[Hz] = 0.00282 t[s] + 2737.2$ for lower level ($t$ time of experiment in sec).

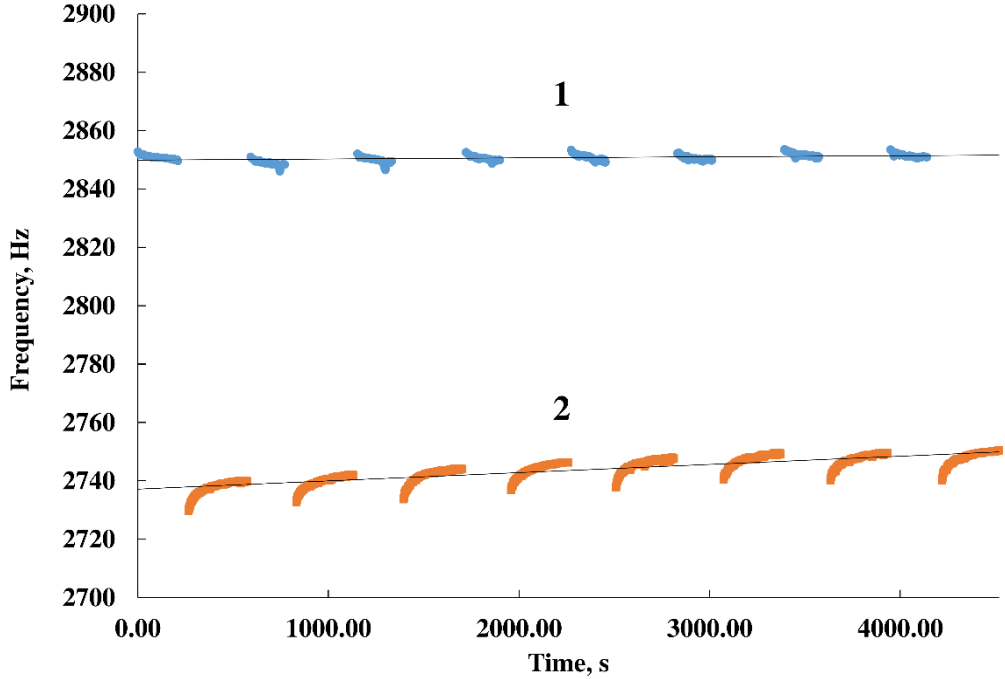

Figure 3. Frequency dependence on time after applying a load of 250 g and subsequently applying a short electrical pulse with amplitude 153 V: 1 - frequency values after wire hardening, 2 - frequency values after applying a load.

The main result of the experiments at this stage is that the measurement of the wire frequency can be used to determine the value of residual changes in the wire structure as a result of inelastic deformations and hardening processes by short electrical pulses.

## 3. Impact of X-ray radiation

To investigate the influence of penetrative radiation on vibrating wire characteristics, we have done an experiment using X-ray apparatus RUP-200-15 with a power exposure dose in air of 50 R/s (Aginian et al., 2004). Devices of this type use X-ray tubes with a tungsten target on the anode. The X-ray radiation generated as a result of the target being bombarded by accelerated electrons is mainly bremsstrahlung with a continuous spectrum up to a maximum energy about to the anode voltage (see, for example, (Ivanov et al. 1989)). The experiment was conducted at anode voltage of 150 keV.

Based on the power of the exposure dose of the X-ray apparatus in air $P_{EX}$, the radiation power density $I$ can be calculated using the following ratio (see (Klyuev, 1992))

$$P_{EX} = I \mu_{AIR} / \rho_{AIR},  \qquad (7)$$

where $\mu_{AIR} / \rho_{AIR}$ is the mass energy absorption coefficient for air. In formula (7), the exposure dose should be expressed in units of J/kg, taking into account: 1 R = 0.00877 Gy in air (formula (19) in (Aginian et al. 2004)), i.e. $P_{EX}$ = 4.39E-01 J/kg/s = 4.39E-04 W/g). We also have (X-Ray Mass Attenuation Coefficients, 2025) $\mu_{AIR} / \rho_{AIR}$ = 1.356E-01 cm$^2$/g (for photons with an energy of 150 keV). As a result, we obtain the radiation power density of the X-ray apparatus $I$ = 3.23E-03 W/cm$^2$. Power falling on the wire: $W_{IN} = I d a_X$, where $d$ is diameter of the wire, $a_X$ is the size of the X-ray spot along the wire.

To calculate the losses of the X-ray beam in a wire made of various materials, we will use the Absorption Calculator (SRS Absorption Calculator, 2025). Transmission coefficient $\varepsilon$ of X-ray through the wire is calculated by equation: $\varepsilon = \exp(-\mu d_{eff})$, where $\mu$ is mass energy absorption coefficient in wire material, $d_{eff} = \pi d / 4$ ($d_{eff}$ found from the condition that the cross-sectional area of the wire $\pi d^2 / 4$ is equal to the area of a rectangle with dimensions $d$ and $d_{eff}$ along the direction of X-ray).

Accordingly, the losses of the X-ray beam in the wire were calculated assuming that all losses are transformed into heat. Table 1 shows the calculations for stainless steel (A316), tungsten, and beryllium bronze. In all cases, the wire diameter is taken equal to 100 μm and the X-ray quantum energy is 150 keV. Thermal power deposited into the wire is denoted as $W_{VWM}$, size of the X-ray beam spot is 3x3 mm².

Table 1. Attenuation of X-rays in various materials.

| Material | stainless steel | tungsten | beryllium bronze |
|---|---|---|---|
| $\rho$, g/cm³ | 7.99 | 1.92×10 | 8.33 |
| $\mu$, cm⁻¹ | 5.20×10⁻¹ | 2.64×10 | 6.80×10⁻¹ |
| $\varepsilon$ | 9.96×10⁻¹ | 8.13×10⁻¹ | 9.95×10⁻¹ |
| $W_{VWM}$, W | 3.95×10⁻⁸ | 1.82×10⁻⁶ | 5.17×10⁻⁸ |

Taking into account the small size of the X-ray spot on the wire, we will calculate the corresponding increase in the temperature of the wire relative to the ambient temperature at wire edges using an approximation of a triangular profile with a maximum overheating $\Delta T$.

The maximum value of the temperature profile $\Delta T_\lambda$, in the case where the heat balance between the power supplied to the wire is determined by heat sink only by heat conduction along the wire, is determined by the formula

$$\Delta T_\lambda = \frac{W_{VWM} L}{4 \lambda S}, \tag{8}$$

where $\lambda$ is coefficient of thermal conductivity.

A similar value for heat sink by convection $\Delta T_\alpha$ is determined by the formula

$$\Delta T_\alpha = \frac{2 W_{VWM}}{\pi \alpha d L}. \tag{9}$$

where $\alpha$ is coefficient of convective losses.

The resulting temperature increase $\Delta T$ is determined by the formula

$$\Delta T = \frac{1}{1/\Delta T_\lambda + 1/\Delta T_\alpha}. \tag{10}$$

The results of these calculations are shown in Table 2. The values from Table 1 were used as the power deposited into wire $W_{VWM}$. In the experiments on the RUP-200-15 setup the VWM with following parameters was used: $L$ = 2.4 cm, $d$ = 90 μm, wire material - beryllium bronze, first harmonic generation.

Table 2. Calculation of $\Delta T$ of the wire from various materials.

| Material | stainless steel | tungsten | beryllium bronze |
|---|---|---|---|
| $\lambda$, W/m/K | 1.63×10 | 2.36×10² | 1.70×10² |
| $\alpha$, W/m2/K | 3.00×10⁻¹ | 3.00×10 | 3.00×10 |
| $\Delta T_\lambda$, K | 1.85×10⁻⁵ | 5.94×10⁻⁵ | 2.32×10⁻⁶ |
| $\Delta T_\alpha$, K | 3.15×10⁻⁸ | 1.46×10⁻⁶ | 4.11×10⁻⁸ |
| $\Delta T$ | 3.14×10⁻⁸ | 1.43×10⁻⁶ | 4.04×10⁻⁸ |

As can be seen from Table 2, the temperature increase of the wire is much less than mK range of VWM temperature resolution (see Introduction) and cannot be detected by measuring the frequency of the vibrating wire.

In the work (Aginian et al. 2004), the dependence of VWM frequency on ambient temperature in the range of 30-45 °C before and after exposing the monitor to X-ray radiation for one hour was also analyzed. The difference in the behavior of the corresponding inclines recorded in the experiment was noted as a result of the impact of X-ray radiation on the structure of the irradiated wire. So that such a conclusion is premature and further experiments are needed with increased X-ray intensity and exposure time on the wire. To increase the overheating temperature of the wire, it is also advisable to conduct experiments in a vacuum, since the main heat sink occurs through convection.

**3. Exposure to proton beam**

To investigate the impact of the proton beam current on the structure of the wire material, a proton beam profiling station based on two vibrating wire monitors was used. For profiling the frequency of the wire oscillations was measured during scanning and followed the temperature changes of the wire as the number of protons penetrating the wire at a given position changed. The station was installed at the proton beam outlet of the C18 cyclotron. At the output flange, the proton energy was 18 MeV. The beamline in experimental hall with installed vibrating wire profiling station is presented in Fig. 4.

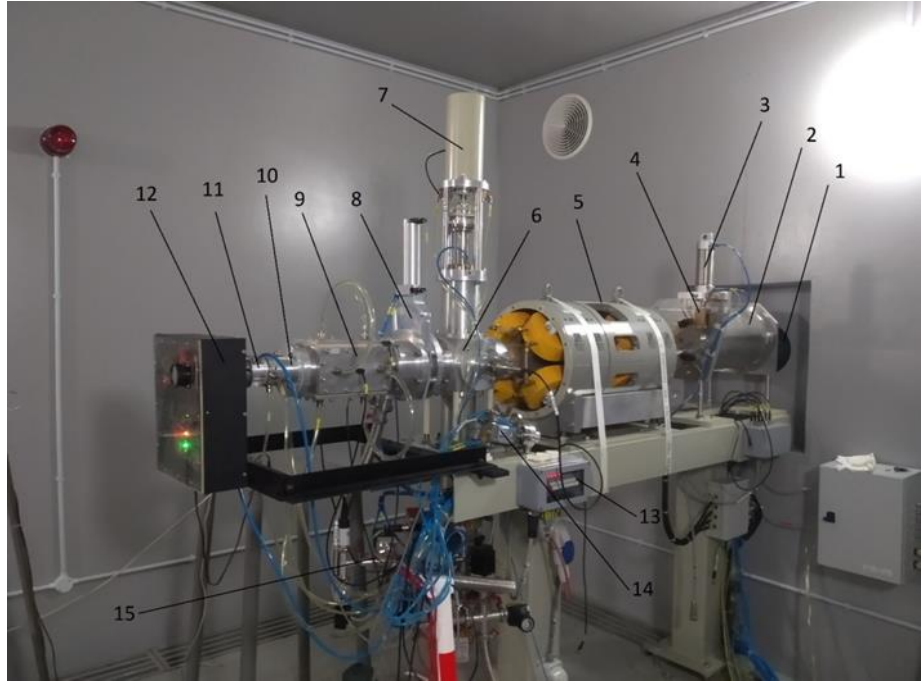

Figure 4. Beamline in experimental hall. 1 - beam injection from the cyclotron hall, 2 - neutron shutter (iron and nickel cylinder for absorption of residual neutrons, open during operation of the beamline, and closed during idle time so that neutrons do not flow from the cyclotron into the experimental hall), 3 - drive mechanism of neutron shutter with pneumatic cylinder, 4 - end switch of the drive mechanism, 5 - a pair of quadrupoles, 6 - viewer for calibration, 7 - fluorescent target viewer for beam position calibration (fed by pneumatic piston), 8 - vacuum valve, 9 - four jaw collimator for beam positioning, 10 - adaptor flange, 11 - vacuum window, 12 - vibrating wire XY profiling station, 13 - electrical contacts with indicators, 14 - solid-state target module, 15 - vacuum tubes with vacuum sensor.

Fig. 5 shows a profiling station consisting of two linear movement systems for vertical and horizontal scanning. VWM parameters: wire from stainless steel (A316), $L$ = 72 mm, $d$ = 100 μm, generation of oscillations on second harmonics.

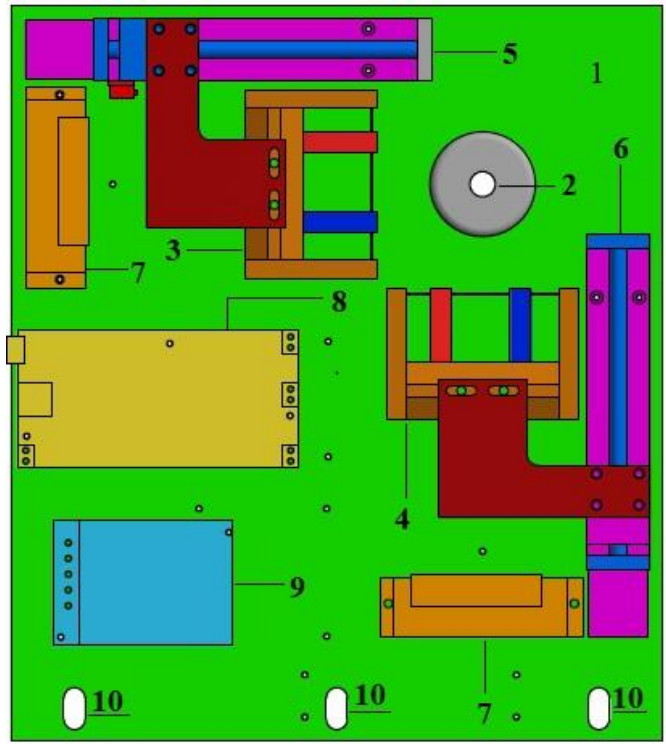

Figure 5. Horizontal and vertical profiling station: 1 - flat support plate with a thickness of 12 mm, on which the station units are located, 2 - collimator with a 12 mm hole, installed on the output flange of the cyclotron beamline, 3 - horizontal beam scanning VWM, 4 - vertical beam scanning VWM, 5 - horizontal VWM feed based on a stepper motor, 6 - vertical VWM feed, 7 - stepper motor drivers, 8 - VWM oscillation autogeneration board with frequency measurement circuit, 9 - power supply, 10 - station fastening holes on the console construction.

In the experiment, the horizontal scanning monitor was set to the position of the profile maximum, while the vertical scanning monitor was placed in the park position and served as a reference for tracking the dependence of the frequency of both monitors on ambient temperature. The proton beam current was set and maintained for a certain period of time (several minutes), and the frequency of the wire oscillations was measured. The process was repeated until the generation of wire oscillations was disrupted. During the experiment, the proton fluence was calculated, since we were interested in the process of accumulation of irreversible changes in the structure of the wire under the action of the proton beam.

At proton beam currents of up to 15 µA, the process of exciting the wire's natural oscillations was not disrupted, and the corresponding frequency served as a measure of the wire's overheating. Fig. 6 shows the combined results of the experiment. The experiment consisted of 31 sessions of proton beam current changes.

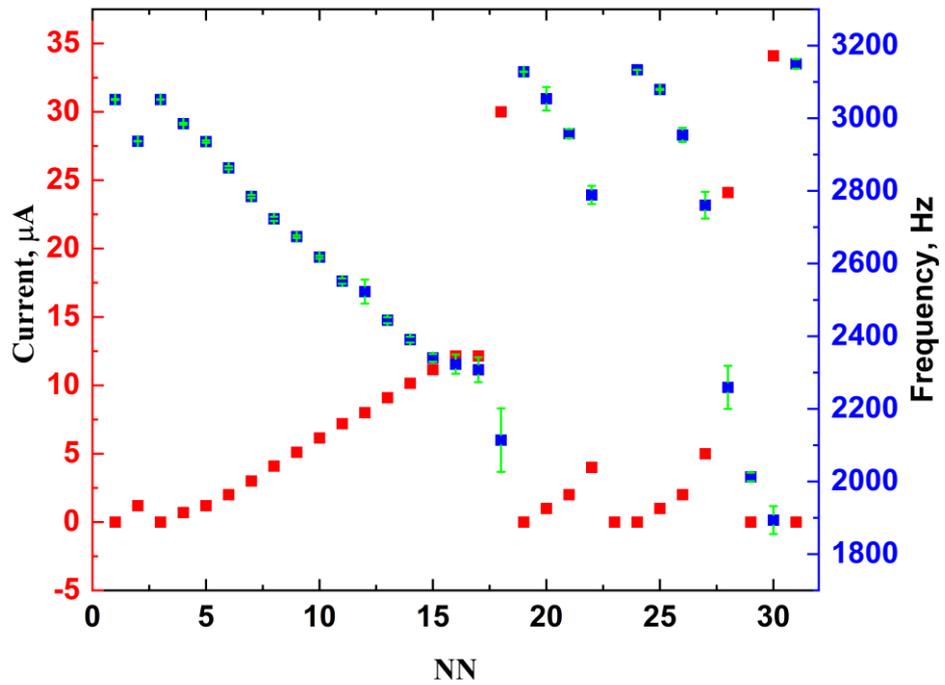

Figure 6. Combined experimental results. The horizontal axis shows the numbers of beam current change sessions. The left vertical axis shows the average proton beam current, and the right vertical axis shows the average frequency of the irradiated wire during a given session.

With an increase in the proton beam current, an increase in the instability of the wire oscillation frequency was observed, apparently due to the instability of the current.

Fig. 7 shows a plot of the relative frequency $w$ (frequency normalized to the value before beam exposure) as a function of the proton beam current. In addition, the value $w$ was corrected for variations in the ambient temperature measured by the second unirradiated monitor: $w = F_1 / F_{10} - F_2 / F_{20}$ ( $F_{1,2}$ are values of frequencies and $F_{10,20}$ initial values of frequencies of the first and second monitors, irradiated VWM is first).

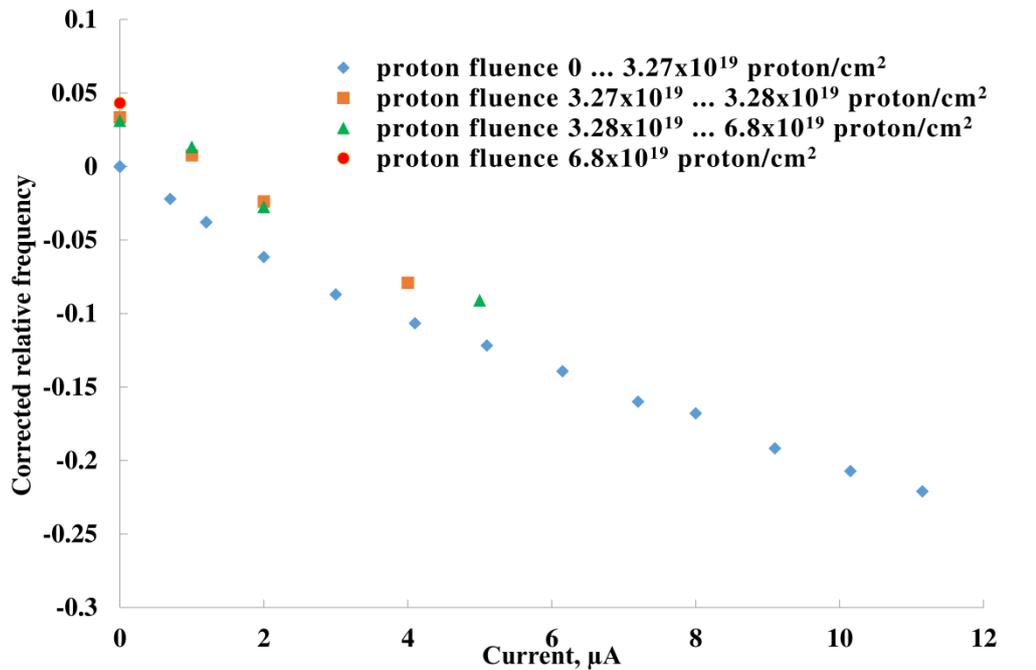

Figure 7. Dependence of the relative frequency $w$ on the proton beam current with correction for the ambient temperature. Blue dots represent sessions in which the proton fluence on the wire increased from zero to $3.27 \times 10^{19}$ proton/cm$^2$, orange dots represent sessions in which it insignificantly increased from $3.27 \times 10^{19}$ proton/cm$^2$ to $3.28 \times 10^{19}$ proton/cm$^2$, and green dots represent sessions in which it increased from $3.28 \times 10^{19}$ proton/cm$^2$ to $6.8 \times 10^{19}$ proton/cm$^2$. The final frequency value with the current turned off when proton fluence reaches $6.8 \times 10^{19}$ proton/cm$^2$ is marked in red.

Table 1 shows the residual changes in the frequency of the wire depending on the proton beam fluence on it when the proton beam was switched off.

Table 1. Residual changes in wire relative frequency depending on proton beam fluence.

| fluence, proton/cm$^2$ | $w$, % |
|---|---|
| 0 | 0 |
| $3.3 \times 10^{19}$ | 3.3 |
| $3.3 \times 10^{19}$ | 3.1 |
| $6.8 \times 10^{19}$ | 4.3 |

As can be seen from Table 1, the residual changes in wire frequency after exposure to a proton fluence $6.8 \times 10^{19}$ proton/cm$^2$ amounted to more than 4%. It should be noted that the stability of the VWM frequency over time intervals of several hours is approximately 0.1 Hz. (Arutunian et al. 2021), which is much less than the frequency shift of 100 Hz recorded in experiments of proton irradiation. Frequency increases from 3051.8 Hz before irradiation to 3149.6 after irradiation with proton fluence $6.8 \times 10^{19}$ proton/cm$^2$. This allows us to confidently assert that a certain hardening of the wire was observed due to structural changes during the irradiation process.

*X-ray structural analysis of an irradiated wire.*

The irradiated wire, as well as the wire from the second monitor of the vibrating wire, which was in the park position, were examined by X-ray diffraction using a Rigaku Miniflex600 X-ray diffractometer (1.5419 Å CuKα radiation). The wire was irradiated only in its central part (distance between magnetic poles - sensor aperture 25 mm, the proton beam size was about 6 mm). Therefore, for the study of irradiated and non-irradiated parts, the wire was divided into three parts: N1-1 (unirradiated), N1-2 (central part, irradiated), N1-3 (unirradiated). The wire from the monitor in the park position is designated as N2 (unirradiated).

For X-ray diffraction measurements, pieces of the wire of the same length were crushed and placed on a glass substrate, with a measurement range of 2θ angles from 10 to 100 degrees. Fig. 8 shows the X-ray diffractogram of sample N1-1

(unirradiated part) in the 2θ angle range from 10 to 100 degrees. The wide peak in the spectrum corresponds to diffraction from the amorphous glass substrate.

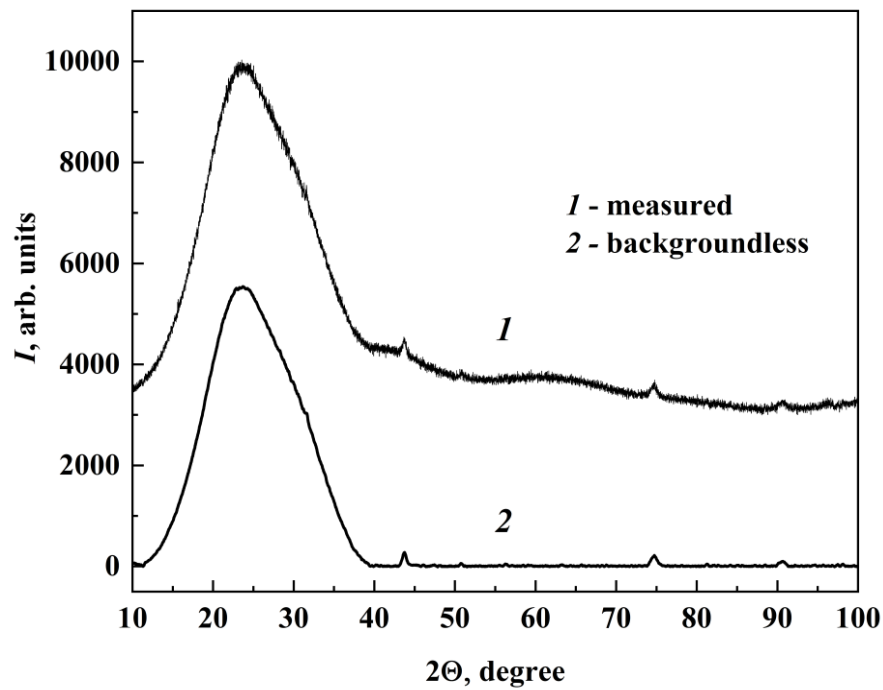

Fig. 8. X-ray diffractogram of the unirradiated part of the wire in the 2θ angle range from 10 to 100 degrees: 1 – actually measured and 2 – processed (without background).

To increase the weak reflections, an X-ray diffractograms were recorded with high amplification in the 2θ angle range from 40 to 100 degrees. Fig. 9 shows the diffractograms of the studied samples of equal mass in the specified range of 2θ angles: N1-1, N1-2 (central part), N1-3, and N2. The same technology was used to prepare powder samples. The figure shows that the lines in the spectra with peaks at 2θ = 43.6, 50.6, 74.6, 90.6, and 95.9 correspond to the (111), (222), (220), (311), and (222) reflections of austenite, the alloy from which the wire is made.

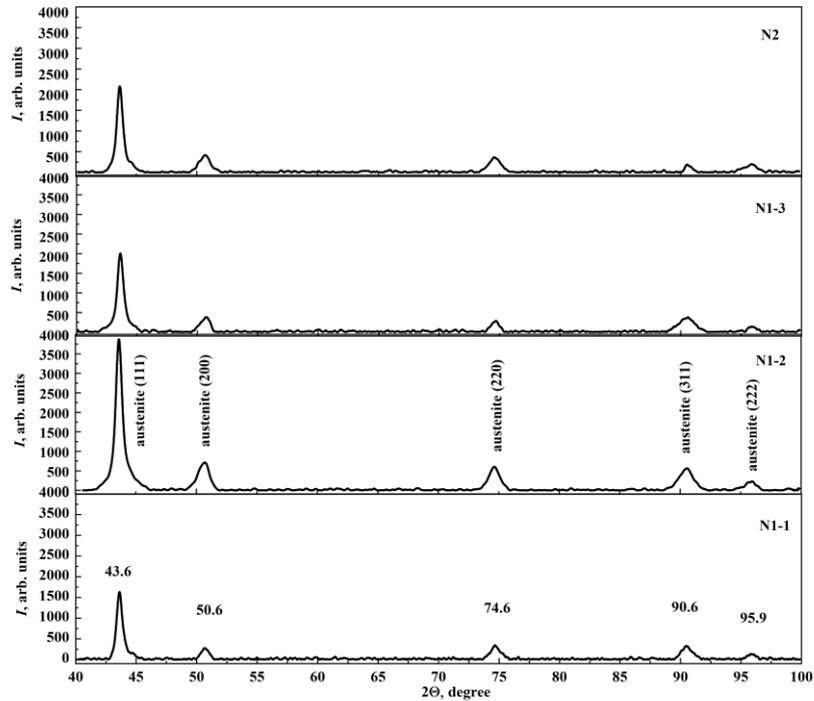

Figure 9. X-ray diffraction spectra of unirradiated samples N1-1, N1-3, and N2 and the irradiated part of the wire N1-2 (central part) in the 2θ angle range from 40 to 100 degrees. Fig. N1-2 shows descriptions of the peaks, and Fig. N1-1 shows the corresponding 2Θ angle values for these peaks.

All graphs are given in the same units. The intensity of all peaks of the irradiated sample (N1-2) increased almost twofold relative to the peaks of the unirradiated samples. Это указывает на изменения в кристаллической структуре материала

This indicates changes in the crystalline structure of the wire material under the influence of radiation within the specified doses (Ermrich, 2011). See also (Barcellini et al., 2019).

## 4. Conclusion

The main purpose of this work is to develop a monitor for structural changes in materials using a vibrating wire. Structural changes in the wire were modelled in several ways. In laboratory conditions, the material was subjected to loads above its elastic limit and rapid thermal heating with quenching in air. Both procedures have a significant and reproducible effect on the natural frequency of the wire, which indicates certain structural changes in the wire material. The parameters of the wire loading and thermal treating procedures are selected so that their cyclic reproduction returns the wire's natural frequency to approximately the same levels. An useful application of the developed procedures could be the frequency tuning of an already stretched wire of VWM. It may also be interesting to investigate the quenching procedure in other atmospheres/vacuum.

The wire was also exposed to the action of ionizing radiation. In the case of X-ray irradiation with intensity about 2000 Roentgen/min and total radiation time 45 min (power density $3 \times 10^{-3}$ W/cm$^2$) (Aginian et al. 2004) no noticeable changes in the frequency of wire oscillations were observed. In the case of irradiation with a proton beam with an energy of 18 MeV and a fluence of up to $6.8 \times 10^{19}$ proton/cm$^2$, a residual change in frequency of more than 4% was observed, which is significantly greater than the frequency stability level in vibrating wire monitors. This allows us to confidently assert that a certain hardening of the wire was observed due to structural changes during the irradiation process. Samples of irradiated and unirradiated sections of the wire were subjected to diffractometric analysis. A significant increase in the intensity of all peaks of the irradiated sample relative to the peaks of the unirradiated samples was revealed. This also indicates changes in the crystalline structure of the wire material under the influence of radiation within the specified doses. Experiments will be continued.

**CRediT authorship contribution statement**

Suren G. Arutunian: Conceptualization, Methodology, Project administration, Supervision, Visualization, Writing – original draft. Narek M. Manukyan: Data curation, Investigation, Validation. Sargis A. Hunanyan: Data curation, Formal analysis, Software, Visualization. Ashot V. Margaryan: Software, Validation, Visualization. Eleonora G. Lazareva: Conceptualization, Data curation, Funding acquisition, Resources, Writing – review & editing. Moses Chung: Conceptualization, Formal analysis, Investigation, Writing – review & editing. Lazar M. Lazarev: Data curation, Formal analysis, Resources. Gevorg S. Harutyunyan: Formal analysis, Software, Visualization. Davit A. Poghosyan: Data curation, Formal analysis, Software. Narek B. Margaryan: Formal analysis, Resources. Natella R. Agamalyan: Conceptualization, Methodology, Writing – review & editing. Manuk N. Nersisyan: Formal analysis, Methodology, Resources.

**Declaration of competing interest**

The authors declare that they have no known competing financial interests or personal relationships that could have appeared to influence the work reported in this paper.

**Data availability**

Data will be made available on request.

**Acknowledgement:**

The authors are grateful to E.R. Arakelova for her assistance in analyzing the diffractograms of the proton beam experiment samples.

The work was carried out with the financial support of the Committee on Science of the Republic of Armenia within the framework of scientific project 25RG-1C107.